# Adaptive Framework for Data Distribution in Wireless Sensor Networks

**Subhabrata Mukherjee[1], Mrinal K. Naskar[2] and Amitava Mukherjee[3]**
[1]Dept of CSE, Jadavpur University, Calcutta 700 032, India
[2]Advanced Digital and Embedded Systems Lab, Dept of ETCE, Jadavpur University, Calcutta 700 032, India
[3] IBM India Pvt Ltd, Salt Lake, Calcutta 700 091, India
amitava.mukherjee@in.ibm.com,subhabrata.mukherjee.ju@gmail.com

**Abstract**—In recent years, the wireless sensor network (WSN) is playing a key role in sensing, collecting and disseminating information in various applications. An important feature associated with WSN is to develop an efficient data distribution and routing scheme to ensure better quality of service (QoS) that reduces the power consumption and the end-to-end data delivery time. In this work, we propose an adaptive framework to transmit data packets from a source to the sink in WSN across multiples paths with strategically distributed data packets so as to minimize the power consumption as well as the end-to-end data delivery time.

## 1. INTRODUCTION

The static sensor network [1]-[3] consists of a large number of smart sensor nodes distributed randomly in a geographically inaccessible area. The minimization of power consumption in WSN is one of the most important design issues because sensor nodes are deployed mostly in geographically inaccessible areas and their energy content thus cannot be easily replenished. However, there are cases like surveillance [1] in battlefield, where the movement of enemy troops is to be monitored continuously. In these cases data must be sensed, processed and transmitted very quickly, so multi-path routing scheme can be used to reduce the data delivery time. Earlier researches [1],[2],[4],[5] showed that multi-path communication can improve the end-to-end data delivery time by taking recourse to simultaneous data transfer over multiple spatial paths.

In this work, we have proposed a method to divide the data into blocks, taking various network factors into consideration, and sending them simultaneously along the different paths available. We have implemented this framework with Mac 802.11 and have shown that it achieves an optimal data delivery time and power consumption when compared to a few other existing protocols as well as the traditional Mac 802.11. The framework we propose considers the network to be a single event model where only one event can be served at a time while others have to wait for their turn. But this lays an essential foundation for a multi-event model where multiple sources can transmit simultaneously which can be obtained by superposition of several single event models (each corresponding to a single source) and leaving the collision resolution responsibility to the underlying MAC layer.

The organization of the rest of the paper is as follows: Section 2 gives a related work. Section 3 describes the system framework for multi-path routing in WSN. Section 4 enhances this framework to introduce an adaptive mechanism for multi-path routing and data distribution. Section 5 gives the simulation results followed by concluding remarks.

## 2. RELATED WORK

There has been a lot of research works in the wireless sensor network area. There have been extensive studies on routing and data distribution in wireless sensor networks. A number of metrics have been used to assess the routing quality, among which the most common and widely used metric has been the hop count. The protocols that use shortest path routing include Dynamic Source Routing (DSR) [6][7], Ad-hoc On-demand Distance Vector routing (AODV) [8], Destination Sequence Distance Vector Routing (DSDV) [9]. In this paper we have taken this as Scheme 1 where all the data has been forwarded via the path with the minimum hop count from the source to the destination. Network reliability can be improved by using multiple paths from the source to the destination instead of using a single path [1][4][5]. In [10], multipath routing is used to increase the reliability of WSNs. The proposed scheme splits up the data into smaller subpackets of equal size and sends them via the multiple paths available. We take this as Scheme 2 where all the data is split up equally and sent across the multiple spatial paths available. We have shown that our proposed framework (Scheme 3) is better than the above 2 schemes as our suggested scheme takes various factors like hop count, energy dissipation due to transmission and reception, bit rate and various other network factors into consideration while distributing the data along multiple spatial paths. [11] suggests an efficient multipath protocol (DCHT) for the wireless sensor network and establishes its efficiency over some other protocols like the Directed Diffusion [12], EDGE [13], C-MFR [14]. We have shown by simulation that the proposed framework used over Mac 802.11 gives a better throughput than DCHT over different network sizes.

## 3. SYSTEM FRAMEWORK

**System Description**

We consider a large number of smart sensors randomly distributed in a geographically hostile area. Any sensor can act as a source node and there is only one sink node (base station). The sink node alone will receive all the data sent by sensors. Each sensor operates on limited battery. Energy is consumed mostly in transmission and receiving data at its radio transceiver. Energy is also consumed when the nodes are sensing or processing data. Each node senses information and delivers it to the sink through a set of paths, each comprising of multiple hops. Each sensor $n \in N$ has a unique identifier. The data sensed by each node is divided among the spatial multi-paths so that the energy consumed in the process and the net end-to-end data delivery time is minimized.

### 3.2 Communication Delay

The data delivery time in any path while sending data from the source to the sink consists of two components:
  a) *Queuing/processing delay*: we consider $q_j$ as the average queuing delay per packet per hop for the jth path.
  b) *Transmission/reception delay*: this delay per packet per hop for the jth path is modeled as $S/b_j + l_j$.

(Where S is the packet size, $b_j$ is the link speed in bits/sec in the jth path and $l_j$ is the link delay in the jth path).

Thus the amount of time required for a data packet to traverse a link (over one hop) along with the queuing delay is defined to be
$$\tau_j = S/b_j + l_j + q_j \qquad (1)$$

Thus, the total delay in the jth path to send $\Delta_j$ data packets over $H_j$ hops is given by
$$T_j = (\Delta_j * \tau_j * H_j) \qquad (2)$$

### 3.3 Energy Consumption

Energy consumption in WSN can be largely categorized into two parts:
  a) *Communication*
  b) *Sensing and processing*

The communication related energy consumption is due to transmission and reception. First we find an expression for the total energy consumed by all the nodes in the jth path.
An energy dissipation model for radio communication similar to [15] [16] has been assumed. As a result the energy required per second for successful transmission of each node ($E_{ts}$) is thus given by,

$$E_{ts} = e_t + e_d d^k \qquad (3)$$

(Where $e_t$ is the energy dissipated in the transmitter electronic circuitry per second to transmit data packets and $e_d d^k$ is the amount of energy required per second to transmit over a distance d and k is the path loss exponent (usually $2.0 \le k \le 4.0$)).

The distance d, must be less than or equal to the radio range $R_{radio}$, which is the maximum inter-nodal distance for successful communication between two nodes. If $T_{1b}$ is the time required to successfully transmit a bit over a distance d then total energy to transmit a bit for each node is

$$E_t = (e_t + e_d d^k) T_{1b} \qquad (4)$$

If $e_r$ is the energy required per second for successful reception and if $T_{2b}$ is the total time required by a sensor to receive a bit then the total energy to receive a bit for each node is

$$E_r = e_r T_{2b} \qquad (5)$$

If we take $d_j$ as the *average inter-hop* distance in the jth path, then $d_j$ can be *approximated* as $d_j = T/H_j$
 (Where T is the total distance between source and sink and $H_j$ is the number of hops in the jth path).
If $\Delta_j$ is the number of packets pushed in the jth path, the total energy dissipation, due to communication, by each node in the jth path is $(E_t + E_r) * \Delta_j * S$.
If $H_j$ is the number of hops in the jth path, the number of nodes in the jth path is given by $H_j + 1$.
Power dissipation at each node due to minimum computation and sensing can be assumed to take place approximately at an effective rate $K_r$. The total power consumed by all the nodes due to sensing and processing is equal to $K_r*$(number of nodes) which is independent of the data division.
Thus the total energy dissipation in the jth path is

$$\begin{aligned} E_j &= [E_t + E_r] * \Delta_j * S * (H_j + 1) + K_r * (H_j + 1) \\ &= [(e_t + e_d * (T/H_j)^k) * T_{1b} + e_r * T_{2b}] * \Delta_j * S * (H_j + 1) + \\ &\quad K_r * (H_j + 1) \qquad (6) \end{aligned}$$

## 4 ADAPTIVE MECHANISM FOR MULTI-PATH ROUTING

### 4.1.1 Creation of the routing table and variable estimation

When a sensor node first joins the network it finds a set of paths to each of the other nodes in the network such that the paths are mutually node exclusive i.e., the nodes in the jth path are distinct from those of the ith path.
During route discovery, the source node broadcasts "hello" packets, which contain the id of the source node and special control information so that other nodes can identify that packet as a special control packet. Each node on receiving a "hello" packet sends a "reply" packet as soon as it can back to the source node which contains its id and other parameters like $K_r$, $H_j$, $e_t$, $e_d$, $T_{1b}$, $T_{2b}$. To get $\tau_j$ of each route, the source node divides the total time between the sending of the "hello" packet and getting a "reply" packet from each node by two. Note that, using "hello" packets we do not require the values of the different components of $\tau_j$ i.e. $b_j$, $l_j$ and $q_j$.
Each node creates a routing table containing the above-mentioned information. Each entry is indexed by a destination node and a set of paths (having the above mentioned characteristics) to reach the destination node and information about those paths as obtained during the route discovery phase.

The sensing event in WSN is assumed to be event driven As soon as a sensor detects an event in its vicinity it checks its routing table. It distributes data among the multipaths obtained from the routing table. The data (D) sensed by a sensor node is thus divided into datasets $\Delta_j$ for j ∈ [1, n], which is distributed over multiple spatial paths, which is done in such a way so as to optimize the end-to-end data delivery time and the power consumption of the network. *The creation of routing table and updation presents an overhead but this is done only once during network setup or when there are multiple node/link failure or multiple new nodes come up.*

### 4.1.2 Three Schemes and Optimization Algorithm

Earlier researches [1], [2], [4], [5] showed that if we send all the data packets from source to sink via a single path (*Scheme 1*) then the end-to-end data delivery time is often more than that obtained when we distribute the data packets equally in all paths and send them simultaneously (*Scheme 2*).

In ideal situations with no congestion, no retransmission and transmission, reception, sensing power of all the nodes being the same in both single and multiple paths, the total power consumption in Scheme 2 is often greater than that in Scheme 1 due to the involvement of more number of sensors in the multiple paths, route discovery mechanisms involving a greater number of sensor nodes, greater cumulative power dissipation due to sensing, tranmission, reception etc.

Our objective is to distribute and route the data packets through several paths available from the routing table, for each source node, in such a way that the end-to-end data delivery time is even less than *Scheme 2* and the total power consumption of the system lies between that of *Scheme 1* and *Scheme 2* but close to that of Scheme *1*. We call this *Scheme 3*.

If we analyze the expression term of $E_j$ we find that $E_j$ decreases as $H_j$ increases which means short hops are favorable. But on the other hand as $H_j$ increases delay also increases due to processing by multiple nodes. Hence we need a trade-off between delay and energy consumption.

We introduce a term here called the energy-delay product (EDP).
The EDP for the jth path is defined as,
$EDP_j = E_j * T_j$
$= [\{(e_t+e_d *(T/ H_j)^k)* T_{1b} + e_r *T_{2b}\}*\Delta_j*S*(H_j +1)+ K_r*(H_j+1)] * (\Delta_j*\tau_j*H_j)$ (7)

Now to reduce the overall data delivery time and the net energy consumption of *Scheme 3*, we seek to make the energy-delay product of every path of Scheme 3 less than or equal to the average energy-delay product of Scheme 2.

The average energy-delay product ($EDP_{avg}$) of Scheme 2 is given by,
$EDP_{avg} = E_{avg} * T_{avg}$
$\approx [\{(e_t+e_d *(T/ H_{avg})^k)*T_{1b} + e_r*T_{2b} \}*(D/n)*S*(H_{avg} +1)+K_r*(H_{avg}+1) ] * ((D/n)*\tau_{avg}*H_{avg})$ (8)
($\Delta_j$=D/n, as data is distributed equally in all the paths in Scheme 2).

Thus to make the net energy-delay product of Scheme 3 less than that of Scheme 2, $EDP_j <= EDP_{avg}$ for j∈ [1, n]. Hence

$[\{( e_t+e_d *(T/ H_j)^k)* T_{1b} +e_r*T_{2b}\}*\Delta_j*S*(H_j +1)+K_r* (H_j+1)]*(\Delta_j*\tau_j*H_j) <= [\{(e_t+e_d*(T/ H_{avg})^k)*T_{1b}+e_r*T_{2b} \} * (D/n) * S * (H_{avg}+1)+K_r*(H_{avg}+1) ] * ((D/n)*\tau_{avg}*H_{avg})$ (9)

Subject to the constraint,
$\sum_j \Delta_j = D$ for j ∈ [1, n] (10)

Equation (9) gives the maximum number of data packets $\Delta_j$ that can be pushed in the jth path keeping the data delivery time and the net energy consumption less than that of *Scheme 2*.

Here one may argue that, by reducing the EDP of the jth path, it may so happen that one component of the EDP may increase and the other may decrease so that the overall EDP for the jth path decreases. So our objective of reducing both the components of EDP i.e. data delivery time and net energy consumption may not be achieved. But this argument does not hold water since the *only variable for the jth path is* $\Delta_j$. The equation above gives a threshold value of $\Delta_j$ that can be pushed in the jth path keeping both energy consumption and delay under a threshold value. Since the data delivery time and the net energy consumption are both an increasing function of $\Delta_j$, for any path if one of them decreases the other is also bound to decrease depending on $\Delta_j$, keeping all other factors constant.

The solution to (9) + (10) will give the value of $\Delta_j$ for j∈ [1, n] i.e., the number of data packets to be sent in the jth path in *Scheme 3*. Generally, the computing resources at a node are limited and the classical optimization problem solving techniques require significant computational resource and time. Here we develop a method that does not require much computational resource or time although it may give a sub optimal solution, which nevertheless achieves our purpose.

Equation (9) is of the form $A\Delta_j^2 + B\Delta_j \leq C$ (where A, B, C are constants). We want to find the maximum value of $\Delta_j$ satisfying the condition above. So, we first solve equation (9) replacing the inequality sign by equality i.e.,

$[\{(e_t+e_d *(T/ H_j)^k)* T_{1b} +e_r*T_{2b} \}*\Delta_j*S*(H_j +1)+K_r*(H_j+1) ]*(\Delta_j*\tau_j*H_j)=[\{(e_t+e_d*(T/H_{avg})^k)*T_{1b}+e_r*T_{2b}\}*(D/n)*S*(H_{avg}+1) +K_r*(H_{avg}+1) ] * ((D/n)*\tau_{avg}*H_{avg})$ (11)

The R.H.S of the equation is a constant as the values of all the terms there are obtained from the routing table. The L.H.S of the equation is only a function of $\Delta_j$ and $H_j$ for j ∈ [1, n].

The above equation (11) is easily solved by substituting values of the constants and the value of $H_j$ *for each of the n spatial paths*.

The data volume $\Delta_j$ for $j \in [1, n]$ obtained from the equation (11) is the maximum number of data packets that can be sent in the particular jth spatial path.

To satisfy constraint (10), the actual number of data packets to be sent into each jth spatial path $j \in [1, n]$ is given by

$$(\Delta_j / \sum_j \Delta_j)*D \quad for\ j \in [1, n] \quad (12)$$

### 4.2 Detection of faulty nodes or links

If there is any packet drop it can be due to collision, node or link failure. In case of a packet drop it needs to be retransmitted. If the 2 sources keep on colliding with each other and the data packets are dropped again and again, then after a maximum number of, say, 'm' attempts the process is aborted.

Now, either the transmitting node 'a' or the receiving node 'b' or the link connecting 'a' or 'b' is at fault. In that case 'a' performs a self-check by transmitting a beacon packet to another neighboring node 'c'.

Case 1: If 'a' fails in this case too then it is at fault. Every node starts a timer from the time it is supposed to receive any data packet from its neighboring node in any transfer round. In case 'b' fails to receive any packet from 'a' within m* $\tau_j$ secs, then it concludes 'a' is at fault.

Case 2: If 'a' succeeds to send the beacon packet to 'c', then either 'b' or the link connecting 'a' and 'b' is at fault.

Node 'b' in Case (1) and node 'a' in Case (2) chooses the redundant node 'd' nearest to it and assigns it the serial number of the node that just failed. The information is communicated to the neighboring nodes of the failed node so that they can update their routing tables. Then the communication proceeds normally. For the remaining of the data transfer round and till the failed node is replaced, the node 'd' performs the functions of node 'a' in Case (1) and node 'b' in Case 2.

### 4.3 Route discovery Algorithm: some important features

a) The source node only initiates route discovery algorithm when it joins the network or there is any change in the network due to failure of multiple nodes or when multiple new nodes join the network.
b) The paths discovered are all node disjoint.
c) The route discovery algorithm gives a set of paths between source and sink along with the information about various nodes and path parameters in each jth path like the values of $K_r$, $H_j$, $e_t$, $e_d$, $T_{1b}$, $T_{2b}$, $\tau_j$

### 4.4 Step by step procedure

The events considered here are non-overlapping in time. In case multiple events occur simultaneously near the vicinity of a node, the events are processed sequentially. Hence while one event is being active, the others have to wait.

*Step 0:* An event occurs near the vicinity of a node. For each event, the following steps are carried out.
*Step 1:* Initialize the set P=Φ, where P is the set of paths for that event in a spatial domain.
*Step 2:* **If** *the routing table is not created or there is any change in the network since the routing table was last updated, then* call the route discovery algorithm to determine
a) Node-disjoint multi-paths. The maximum number of multi-paths will be less than or equal to the number of nodes within the transmission range of the source node as we consider only node disjoint multi-paths
b) Parameters $K_r$, $H_j$, $e_t$, $e_d$, $T_{1b}$, $T_{2b}$, $\tau_j$ for each jth path
c) Find the average hop count $H_{avg}$.
d) In a certain period of time (in this paper, 'certain period of time' or 'an interval' will mean the time required for successful transmission of data sensed by a particular sensor to the sink) the energy of each node of the network taking part in transmission/reception decreases by $(E_t + E_r)*\Delta_j*S + K_r$ in the jth path. The power of any node not taking part in transmission will decrease by $K_r$ where $K_r$ denotes the effective rate (statistically averaged) of power loss of each node due to sensing.

**Else** consult the routing table to obtain the above mentioned information.
*Step 3:* Solve equation (9) + (10) with the set of input values determined by the route discovery algorithm
*Step 4:* End

## 5  SIMULATION RESULTS

### 5.1 Comparison of Scheme 3 with Scheme 1 and Scheme 2 in terms of Delay and Power Consumption

In order to study the performance of our proposed framework we have run simulation program in NS2. We have first run the route discovery algorithm. We consider an area of 501*501 square meters where 1000 nodes are deployed randomly. The nodes have a transmission radius of 2.4 meters. The simulation has been done with 100 and 200 data packets (size of each data packet is 1 kb). The simulation has been done for multiple runs and the mean result for each scheme has been shown. In this simulation, the routing algorithm gives 5 possible paths between the source node and the sink along with the following parameters-

| | |
|---|---|
| *Hop Counts* | 9,22,5,20,7 respectively |
| *Initial energy of each node* | 23760 joules for all (calculated on basis of MICA 2 motes) |
| *Mac* | Mac/802.11 |
| *Bit rate* | 50 kbps |
| $K_r$ | 0.024 Watts (assuming 40% duty |
| | cycle for mica motes) |
| *Transmission Power* per packet per hop | 1024 µJ/sec |
| *Receiving Power* per packet per hop | 819.2 µJ/sec |
| *Idle Power* | 409.6µJ/s |

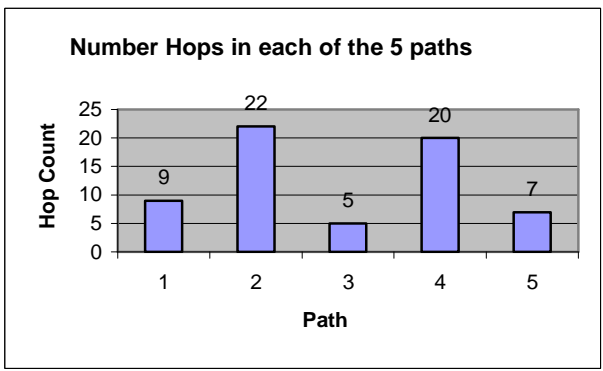

**Figure 1: Hop Count in different paths selected by Dijkstra's algorithm**

With the above parameters as input, the optimization algorithm divides data over each selected path. Figure 1 gives the number of hop counts in each path. Figure 2 gives the energy consumption in joules in the 3 Schemes i.e., when the data packets are routed via a single path (Scheme 1), when the data packets are equally distributed among 5 paths (Scheme 2) and when the data packets are routed using our proposed Scheme (Scheme 3). The energy comparisons in figure 2 are done with 100 and 200 data packets. We see that the energy consumption in our Scheme lies between that of single path and multipath equi-distribution Scheme but closer to the single path routing Scheme.

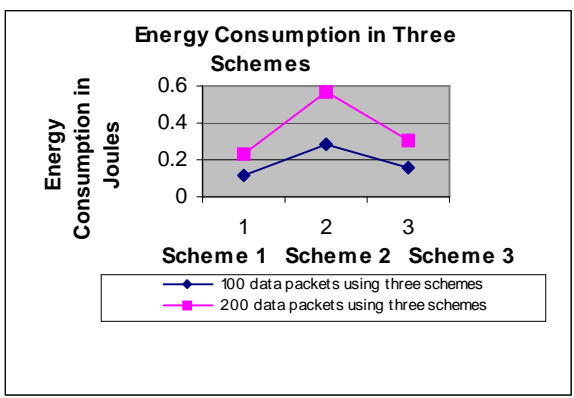

**Figure 2: Comparison of Energy Consumption in the 3 Schemes**

The Table 1 gives the data delivery time for individual paths using Scheme 2 and Scheme 3 for the 2 categories of simulation i.e., with 100 and 200 data packets.

Figure 3 compares the data delivery time in the three Schemes. The Scheme 1 causes maximum data delivery time while our proposed Scheme 3 gives minimum data delivery time. The data delivery time for Scheme 2 is less than than Scheme1 for using multiple paths. The data delivery time in Scheme 3 is even less than Scheme 2, as it takes various network factors in consideration during data distribution which Scheme 2 does not do. The data delivery time in Scheme 2 is in between the two. Figure 4 gives the data distribution in each path when simulation is done with 100 and 200 data packets in Scheme 3.

| | Simulation using | | | |
|---|---|---|---|---|
| Path | 100 data packets | | 200 data packets | |
| | Scheme 2 | Scheme 3 | Scheme 2 | Scheme 3 |
| 1 | 3.595 | 3.594 | 7.194 | 7.194 |
| 2 | 8.794 | 3.514 | 17.59 | 7.034 |
| 3 | 1.994 | 3.694 | 3.994 | 7.394 |
| 4 | 7.994 | 3.594 | 15.994 | 7.194 |
| 5 | 2.794 | 3.634 | 5.594 | 7.274 |

**Table 1: Delay in the individual path (in secs) in Scheme 2 and Scheme 3 when simulation is done with 100 and 200 data packets**

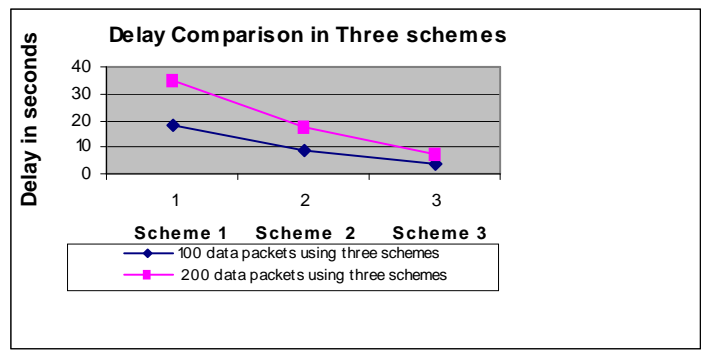

**Figure 3: Delay comparison in the three schemes**

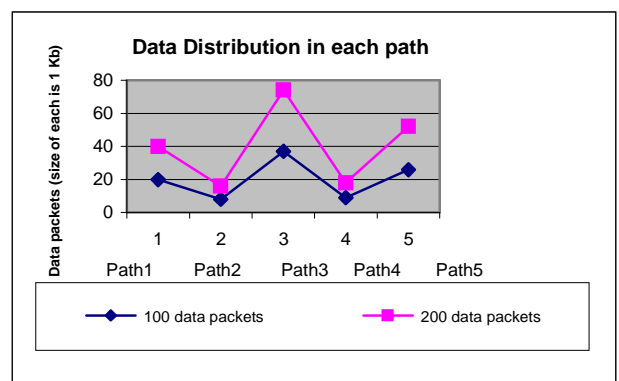

**Figure 4: Data Distribution in the five paths in Scheme 3**

From the above results we have observed that the proposed framework is the best of the three Schemes where the net energy consumption and the end-to-end delay both are minimized.

### 5.2 Comparison of a few other routing protocols with the suggested protocol

[11] gives an efficient multipath protocol (DCHT) for the wireless sensor network and establishes its efficiency over some other protocols like the Directed Diffusion [12],EDGE [13],C-MFR [14]. We have shown by simulation that our suggested framework over 802.11 gives better throughput than DCHT over different network sizes. Hence from the above comparisons, it is apparent that our suggested framework performs better than the before mentioned protocols also. Furthermore, there are no packets losses in the suggested scheme unlike the DCHT protocol where there are packet losses. Thus our scheme is more reliable. As the network size increases the throughput falls as the average path length increases.

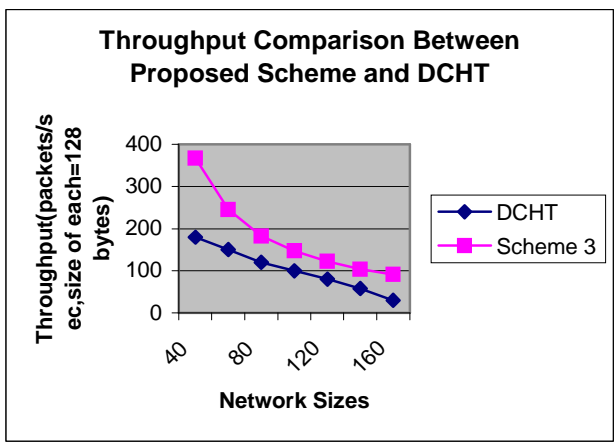

**Figure 5. Throughput of DCHT, Proposed Scheme with different network sizes**

### 6 CONCLUSION AND FUTURE WORK

From the above observations we can conclude that the proposed framework serves well because it is adaptive and computes the data distribution based on various network factors. It achieves its objective of both power and delay optimization. It is energy-aware, delay-aware, maintains an uniform load distribution. Uniformity and reliability is ensured by including nodes in a path based on a number of factors like hop-count, residual energy and distributes data along the paths accordingly to make sure no node is overwhelmed with data. This framework can be implemented with any existing MAC (it has been tested with 802.11,802.15.4) and improves its performance considerably. This proves to be one of the strongest features of the proposed framework. The events considered here are non-overlapping in time or sequential. When multiple events occur simultaneously in the vicinity of a node or when multiple sources transmit at the same time, congestion in the network increases resulting in the increase in collisions and dropped data packets. A possible solution is to superimpose various single-event models so that each source acts independently, irrespective of the other sources, and the paths will be *locally node-disjoint* for any source and not *globally node-disjoint*. We can leave the collision and congestion control for the underlying MAC layer to handle.

**REFERENCES**

**1**. T, He, S. Krishnamurthy, J. A. Stankovic, T. Abdelzaher, L. Luo, R. Stoleru, T. Yan and L. Gu, "Energy-Efficient Surveillance System Using Wireless Sensor Networks", Proceeding of ACM Mobisys, 2004, pp 06-09

**2**. W. R Heinzelman, A. Chandrakasan and H. Balakrishnan, "Energy efficient Communication protocol for wireless micro sensor networks", IEEE Proc Hawaii International Conf Sys, Jan 2000

**3**. S. Sarkar, A. Bhattacharya, M. K. Naskar, A. Mukherjee "An Efficient Multi-path Routing Protocol for Energy Optimization in Wireless", Accepted to SNCNW 2008, Apr, Sweden.

**4**. S. K. Das, A. Mukherjee, S. Bandopadhyay, D. Saha and K. Paul, "An adaptive framework for Qos routing through multiple paths in ad hoc wireless networks, Journal of Parallel and Distributed Computing, Vol 63, 2003, pp 140-153.

**5**. R. Banner and A. Orda, "Multipath Routing Algorithms for Congestion Minimization", IEEE/ACM Transactions on Networking, Vol. 15, No. 2, Apr 2007, pp 413-424.

**6**. David B Johnson and David A Maltz, Dynamic source routing in ad hoc wireless networks," in Mobile Computing, Imielin-ski and Korth, Eds., vol. 353, pp. 153-181. Kluwer Academic Publishers, 1996

**7**. D. B. Johnson and D. A. Maltz, Dynamic source routing in ad hoc wireless networks, Mobile Computing, vol. 353, 1996.

**8**. Charles E. Perkins and Elizabeth M. Royer, Ad hoc on-demand distance vector routing," in Proc. of the 2nd IEEE Workshop on Mobile Computing Systems and Applications, 1999, pp. 90-100.

**9**. C. E. Perkins and P. Bhagwat, Highly dynamic destination-sequenced distance-vector routing (dsdv) for mobile computers, in SIGCOMM '94: Proceedings of the conference on Communications architectures, protocols and applications. New York,NY, USA: ACM, 1994,pp. 234 - 244.

**10**. S. Dulman, T. Nieberg, J. Wu, P. Havinga, "Tradeoff between Traffi Overhead and Reliability in Multipath Routing for Wireless Sensor Networks",WCNC Workshop, New Orleans, Louisiana, USA, March 2003.

**11**. Shuang Li, Raghu Kisore Neelisetti,Cong Liu and Alvin Lim,"Efficient Multi-Path Protocol For Wireless Sensor Networks",International Journal of Wireless and Mobile Networks (IJWMN),Vol.2,No.1,February 2010

**12**. C. Intanagonwiwat, R. Govindan, D. Estrin, J. Heidemann, and F. Silva, Directed diffusion for wireless sensor networking, IEEE/ACM Transactions on Networks, vol. 11,no. 1, pp. 2-16,2003.

**13**. S. Li, A. Lim, S. Kulkarni, and C. Liu, Edge: A routing

algorithm for maximizing throughput and minimizing delay in wireless sensor networks, Military Communications Conference,MILCOM 2007, IEEE, pp. 1-7, 29-31 Oct. 2007.

**14**. X. Lin and I. Stojmenovic, Location-based localized alternate, disjoint and multi-path routing algorithms for wireless networks, J. Parallel Distrib. Comput., 2002.

**15**. S. Lindsey, C.S. Raghavendra, "PEGASIS: Power Efficient Gathering in Sensor Information Systems", In Proceedings of IEEE ICC 2001, pp. 1125-1130, June 2001.

**16**. Q. Gao, K.J. Blow, D.J. Holding, I.W. Marshall and X.H. Peng, "Radio Range Adjustment for Energy Efficient Wireless Sensor Networks", Ad-hoc Networks Journal, Elsevier Science, January 2006, vol. 4, issue 1, pp.75-82.